\pdfoutput=1
\RequirePackage{ifpdf}
\ifpdf 
\documentclass[pdftex]{sigma}
\else
\documentclass{sigma}
\fi

\numberwithin{equation}{section}

\catcode`@=11
\catcode`@=12

\def\dderpar#1#2#3{\mathchoice%
{{\partial^2 #1\over\partial #2\,\partial #3}}{{\partial^2 #1/\partial #2\,\partial #3}}%
{{\partial^2 #1\over\partial #2\,\partial #3}}{{\partial^2 #1/\partial #2\,\partial #3}}}
\def\derpar#1#2{\mathchoice%
{{\partial#1\over\partial#2}}{{\partial#1/\partial#2}}{{\partial#1\over\partial#2}}{{\partial#1/\partial#2}}}
\newcommand{\ov}[1]{\overline #1}

\begin{document}

\allowdisplaybreaks

\renewcommand{\thefootnote}{$\star$}

\renewcommand{\PaperNumber}{011}

\FirstPageHeading

\ShortArticleName{Symmetries of the Free Schr\"odinger Equation in the Non-Commutative Plane}

\ArticleName{Symmetries of the Free Schr\"odinger Equation\\
in the Non-Commutative Plane\footnote{This paper is a~contribution to the Special Issue on Deformations of
Space-Time and its Symmetries.
The full collection is available at \href{http://www.emis.de/journals/SIGMA/space-time.html}
{http://www.emis.de/journals/SIGMA/space-time.html}}}

\Author{Carles BATLLE~$^\dag$, Joaquim GOMIS~$^\ddag$ and Kiyoshi KAMIMURA~$^\S$}

\AuthorNameForHeading{C.~Batlle, J.~Gomis and K.~Kamimura}

\Address{$^\dag$~Departament de Matem\`atica Aplicada 4 and Institut d'Organitzaci\'o i Control,\\
\hphantom{$^\dag$}~Universitat Polit\`ecnica de Catalunya - BarcelonaTech, EPSEVG, Av.~V.~Balaguer~1,\\
\hphantom{$^\dag$}~08800 Vilanova i la Geltr\'u, Spain}
\EmailD{\href{mailto:carles.batlle@upc.edu}{carles.batlle@upc.edu}}

\Address{$^\ddag$~Departament d'Estructura i Constituents de la Mat\`eria and Institut de Ci\`encies del Cosmos,\\
\hphantom{$^\ddag$}~Universitat de Barcelona, Diagonal 647, 08028 Barcelona, Spain}
\EmailD{\href{mailto:gomis@ecm.ub.es}{gomis@ecm.ub.es}}

\Address{$^\S$~Department of Physics, Toho University, Funabashi, Chiba 274-8510, Japan}
\EmailD{\href{mailto:kamimura@ph.sci.toho-u.ac.jp}{kamimura@ph.sci.toho-u.ac.jp}}

\ArticleDates{Received August 29, 2013, in f\/inal form January 29, 2014; Published online February 08, 2014}

\Abstract{We study all the symmetries of the {free} Schr\"odinger equation in the non-com\-mu\-tative plane.
These symmetry transformations form an inf\/inite-dimensional Weyl algebra that appears naturally from
a~two-dimensional Heisenberg algebra generated by Galilean boosts and momenta.
These inf\/inite high symmetries could be useful for constructing non-relativistic interacting higher spin
theories.
A~f\/inite-dimensional subalgebra is given by the Schr\"odinger algebra which, besides the Galilei
generators, contains also the dilatation and the expansion.
We consider the quantization of the symmetry generators in both the reduced and extended phase spaces, and
discuss the relation between both approaches.}


\Classification{81R60; 81S05; 83C65}

\vspace{-2mm}

\renewcommand{\thefootnote}{\arabic{footnote}}
\setcounter{footnote}{0}

\section{Introduction and results}

The symmetries of a~free massive non-relativistic particle and the associated Schr\"odinger equation have
been investigated.
The projective symmetries of the Schr\"odinger equation induced by the transformation on the coordinates
$(t,\vec x)$ are well known.
They form the Schr\"odinger group~\cite{Hagen:1972pd, Jackiw:1972cb,Kastrup:1968zza,Niederer:1972zz} that,
apart from the Galilei symmetries, contains the dilatation and the expansion.
Recently Valenzuela~\cite{Valenzuela:2009gu} (see also~\cite{Bekaert:2011qd}) discussed higher-order
symmetries of the free Schr\"odinger equation.
These symmetry transformations form an inf\/inite-dimensional Weyl algebra constructed from the generators
of space-translation and the ordinary commuting Galilean boost.
The extra symmetries that do not belong to the Schr\"odinger group correspond to higher spin symmetries.
These transformations are not induced by the transformations on the coordinates but they map solutions into
solutions of the Schr\"odinger equation.

In the case of $2+1$ dimensions, the Galilei group admits two central
extensions~\cite{BGO,BGGK,HMS03,HMS10, LL}, one associated to the exotic non-commuting boost and other
appearing in the commutator of the ordinary boost and spatial translations.
The non-relativistic particle in the non-commutative plane was introduced in~\cite{Lukierski:1996br} by
considering a~higher order Galilean invariant Lagrangian for the coordinates $(t,\vec x)$ of the particle.
A f\/irst order Lagrangian depending on the coordinates $(t,\vec x)$ and extra coordinates $\vec v$ was
introduced in~\cite{Duval:2001hu}.
For these Lagrangians there are two possible realizations, one with non-commuting (exotic) boosts, and the
other with ordinary commuting boosts~\cite{BGGK,Horvathy:2004fw} (see~\cite{HMS10} for a~review).

In this paper we study all the inf\/initesimal Noether symmetries of a~massive free particle in the
$(2+1)$-dimensional non-commutative plane.
The Noether symmetries are constructed from the Heisenberg algebra of commuting boosts $X_i$ and the
generators of translations~$P_i$, $\{X_i,P_j\}=\delta_{ij}$, $i,j=1,2$, all of which are constants of
motion { and are written explicitly in terms of the initial conditions}.
The algebra of these symmetries is the inf\/inite-dimensional Weyl algebra associated with the Heisenberg
algebra.
A general element of the Weyl algebra is given by~$\mathfrak{G}(X_i, P_j)$.
The generators given by higher degree polynomials do not form a~closed algebra for any f\/inite degree.
These inf\/inite symmetries are the non-relativistic counterpart of all the symmetries of the free massless
Klein--Gordon equation~\cite{Eastwood:2002su}.
There is no known realization of this Weyl algebra for an Schr\"odinger equation with interaction.
These symmetries could be useful to construct a~non-relativistic analogue of Vasiliev's higher spin
theories~\cite{Vasiliev:2004cp}.

\looseness=-1
The subset of generators constructed up to quadratic polynomials of $ (X_i,P_j)$ form a~f\/inite-dimensional sub-algebra, which in turn contains the 9-dimensional Schr\"odinger algebra.
We study the realization of this algebra in the classical unreduced phase-space, as well as in the reduced
one, the later appearing due to the presence of second class constraints.
We also study all the symmetries of the free Schr\"odinger equation in the non-commutative plane.
The symmetries are in one to one correspondence with the Noether symmetries of the free particle in the
non-commutative plane.
This analysis is done in the quantum reduced phase space, as well as in the extended one.
In the extended space we impose non-hermitian combinations of the second class constraints.
In this case we consider two representations for the physical states, namely a~Fock
representation~\cite{Horvathy:2004fw} and a~coordinate representation.
We study the Schr\"odinger subalgebra in detail, and we show the equivalence between the reduced and
extended space formulations.
We show that, in general, the quadratic (and higher) generators in the extended space contain second order
derivatives and hence do not generate point transformations for the coordinates.

The organization of the paper is as follows.
In Section~\ref{section2} we construct all Noether symmetries of the massive particle in the
non-commutative plane.
Section~\ref{section3} is devoted to the study of the quantum symmetries of the Schr\"odinger
equation.

\vspace{-1mm}

\section{Classical symmetries of the non-relativistic particle\\ Lagrangian in the non-commutative plane}
\label{section2}

In this section we introduce a~f\/irst order Lagrangian describing a~particle in the non-commutative
plane~\cite{Duval:2001hu}, and present the corresponding Hamiltonian formalism.
The main result of the section is the construction of all the Noether symmetries of the non-relativistic
particle in the non-commutative plane (equations~\eqref{XPdef}--\eqref{variationL} and the ensuing
discussion).
For the sake of completeness, we review the construction of the standard and exotic Galilei algebras and of
the Sch\"odinger generators~\cite{Banerjee:2005zt, BGGK,HMS10,Horvathy:2004fw,Horvathy:2005wf}.
We also perform the reduction of the second class constraints of the system for later use in the
quantization in the reduced phase space.

The f\/irst order Lagrangian of a~non-relativistic particle in the non-commutative plane, see for
example~\cite{Duval:2001hu}, is given by
\begin{gather}
\label{action}
{\cal L}_{\rm nc}=m\left(v_i\dot x_i-\frac{v_i^2}{2}\right)+\frac{\kappa}{2}\epsilon_{ij}v_i\dot v_j,
\qquad
i,j=1,2,
\end{gather}
where the overdot means derivative with respect to ``time'' $t$.
This Lagrangian can be obtained using the NLR method~\cite{Coleman1, Coleman2} applied to the exotic Galilei
group in $2+1$ dimensions\footnote{Note that this Lagrangian is not dynamically equivalent to the higher
order Lagrangian for a~non-relativistic particle proposed in~\cite{Lukierski:1996br}.
It can be obtained from~\eqref{action} using the inverse Higgs mechanism~\cite{Ivanov:1975zq}.};
see~\cite{Alvarez:2007fw} for the case of exotic Newton--Hooke whose f\/lat limit gives~\eqref{action}.
The coordinates $x_i$'s are the Goldstone bosons of the transverse translations and $v_i$'s are the
Goldstone bosons of the broken boost.
The $v_i$'s and $\kappa$ are dimensionless.

The Lagrangian~\eqref{action} gives two primary second class constraints
\begin{gather}
\Pi_i=\pi_i+\frac{\kappa}{2}\epsilon_{ij}v_j\approx0,
\qquad
V_i=p_i-mv_i\approx0,
\label{constNon2}
\end{gather}
where $p_i$ and $\pi_i$ are the momenta canonically conjugate to $x_i$ and $v_i$.
The constraints~\eqref{constNon2} satisfy relations
\begin{gather*}
\{\Pi_i,\Pi_j\}=\kappa\epsilon_{ij},
\qquad
\{V_i,V_j\}=0,
\qquad
\{\Pi_i,V_j\}=m\delta_{ij},
\end{gather*}
and the Dirac Hamiltonian is
\begin{gather}
\label{quadratichamiltonian}
H=\frac{p_i^2}{2m},
\end{gather}
up to quadratic terms in the constraints.

From the canonical pairs $(x,v,p,\pi)$ we can get a~new set of canonical pairs $(\tilde x,\tilde v,\tilde p,\tilde\pi)$
given~by
\begin{gather}
\label{canonicaltransformation}
\begin{pmatrix}
\tilde x\\ \tilde p\\
\tilde v\\ \tilde\pi
\end{pmatrix}
=
\begin{pmatrix}
1&-\dfrac{\kappa}{2m^2}\epsilon&\dfrac{\kappa}{2m}\epsilon&-\dfrac{1}{m}\vspace{1mm}\\
&1&&\vspace{1mm}\\
&-\dfrac{1}{m}
&1&\vspace{1mm}\\
&\dfrac{\kappa}{2m}\epsilon&&1
\end{pmatrix}
\begin{pmatrix}
x\\ p\\ v\\
\pi
\end{pmatrix}
.
\end{gather}
In terms of the new variables the constraints~\eqref{constNon2} become a~canonical pair,
\begin{gather}
\tilde v_i=-\frac1m V_i\approx0,
\qquad
\tilde\pi_i=\Pi_i+\frac{\kappa}{2m}\epsilon_{ij}V_j\approx0.
\qquad
\label{constsec0}
\end{gather}
The position and momentum of the particle are expressed as
\begin{gather}
x_i=\tilde x_i-\frac{\kappa}{2m^2}\epsilon_{ij}\tilde p_j-\frac{\kappa}{2m}\epsilon_{ij}\tilde v_j+\frac{1}{m}\tilde\pi_i,
\qquad
p_i=\tilde p_i,
\label{constsecx}
\end{gather}
and the Dirac Hamiltonian~\eqref{quadratichamiltonian} is written as
\begin{gather}
H=\frac{1}{2m} {\tilde p_i}^2.
\label{Ham0}
\end{gather}
The phase space is a~direct product of two spaces.
One is spanned by $(\tilde v,\tilde\pi)$ with the constraints~\eqref{constsec0}
\begin{gather}
\tilde v_i\approx0,
\qquad
\tilde\pi_i\approx0
\label{constsec}
\end{gather}
and thus classically trivial.
The other one is spanned by $(\tilde x,\tilde p)$ with the Hamiltonian~\eqref{Ham0}.
It is a~system of a~free non-relativistic particle in $2+1$ dimensions but with the coordinates~$\tilde x_i$.
In the classical reduced phase space def\/ined by the second class constraints~\eqref{constsec} the
coordinates~$x_i$ become non-commutative (see also Subsection~\ref{section2.1}),
\begin{gather}
\label{NCxixi}
\{x_i,x_j\}^*=\left\{\tilde x_i-\frac{\kappa}{2m^2}\epsilon_{ik}\tilde p_k,\tilde x_j-\frac{\kappa}{2m^2}\epsilon_{j{\ell}}
\tilde p_{\ell}\right\}=\frac{\kappa}{m^2}\epsilon_{ij}.
\end{gather}

If we consider a~point transformation $(x,v)\to(y,u)$
\begin{gather}
\label{def_comm-coor1}
y_i=x_i+\frac{\kappa}{2m}\epsilon_{ij}v_j,
\qquad
u_i=v_i,
\end{gather}
in the Lagrangian~\eqref{action} it becomes
\begin{gather*}
{\cal L}=m\left(u_i\dot y_i-\frac{u_i^2}{2}\right),
\end{gather*}
which is the Lagrangian of a~free non-relativistic particle with the commutative coordinates $y_i$.
Although it has a~form of free particle we keep $x_i$ as the ``position coordinates" of this system.
Local interactions would be introduced at the position~$x_i$ rather than~$y_i$.
The coordinates~$y_i$ in~\eqref{def_comm-coor1} are identif\/ied with the commuting coordinates~$\tilde x_i$
in~\eqref{constsecx}, while $x_i$ are non-commutative as in~\eqref{NCxixi}.

All the Noether symmetries are generated by constants of motion which are arbitrary functions
$\mathfrak{G}(X_i, P_j)$ of
\begin{gather}
X_i=\tilde x_i(0)=\tilde x_i(t)-\frac tm\tilde p_i(t)
\qquad
\text{and}
\qquad
P_i=\tilde p_i(0)=\tilde p_i(t),
\label{XPdef}
\end{gather}
verifying
\begin{gather*}
\{{P}_i,{P}_j\}=0,
\qquad
\{{X}_i,{P}_j\}=\delta_{ij},
\qquad
\{{X}_i,{X}_j\}=0.
\end{gather*}
The Lagrangian~\eqref{action} is quasi-invariant under the transformation generated by $\mathfrak{G}(X_i,
P_j)$.
The canonical variations of $(x,v)$ are
\begin{gather}
\delta x_i=\derpar{\mathfrak{G}}{p_i}=\derpar{\mathfrak{G}}{P_i}-\frac{t}m\derpar{\mathfrak{G}}{X_i}
+\frac{\kappa}{2m^2}\epsilon_{ij}\derpar{\mathfrak{G}}{X_j},
\qquad
\delta v_i=\derpar{\mathfrak{G}}{\pi_i}=-\frac{1}{m}\derpar{\mathfrak{G}}{X_i}.
\label{delxvF}
\end{gather}
When computing the variation of the Lagrangian~\eqref{action} under~\eqref{delxvF}, the $(p_i,\pi_i)$ are
replaced, using the def\/inition of momenta~\eqref{constNon2}, by
\begin{gather}
p_i\to mv_i,
\qquad
\pi_i\to-\frac\kappa2\epsilon_{ij}v_j,
\qquad
X_i\to x_i-tv_i+\frac{\kappa}{2m}\epsilon_{ij}v_j.
\end{gather}
It follows that the variation of the Lagrangian becomes a~total derivative,
\begin{gather}
\delta{\cal L}_{\rm nc}=\frac{{\rm d}}{{\rm d}\tau}\mathfrak{F}(x,v,t),\nonumber
\\
\mathfrak{F}(x,v,t)=[p_i\delta x_i+\pi_i\delta v_i-\mathfrak{G}
]_{p_i=mv_i,\, \pi_i=-\frac\kappa2\epsilon_{ij}v_j}\nonumber
\\
\phantom{\mathfrak{F}(x,v,t)}
=\left[mv_i\left(\derpar{\mathfrak{G}}{P_i}-\frac{t}m\derpar{\mathfrak{G}}{X_i}\right)-\mathfrak{G}
\right]_{p_i=mv_i,\,\pi_i=-\frac\kappa2\epsilon_{ij}v_j}.
\label{variationL}
\end{gather}

All these Noether symmetries generate an inf\/inite-dimensional Weyl algebra.
The Weyl algebra, denoted by $[\mathfrak{h}_2^*]$, can be def\/ined~\cite{Valenzuela:2009gu} as the one
generated by (the Weyl ordered) polynomials of the Heisenberg algebra generators, $(X_i,P_i)$, that we
indicate by $\mathfrak{G}(X_i, P_j)$.
{}$[\mathfrak{h}_2^*]$ is the inf\/inite-dimensional algebra of a~particle in the non-commutative plane.
These inf\/inite symmetries are the non-relativistic counterpart of the complete set of symmetries of the
free massless Klein--Gordon equation~\cite{Eastwood:2002su}.
The existence of a~realization of this Weyl algebra for an interacting Schr\"odinger equation is an
interesting open question.

There are f\/inite-dimensional subalgebras of the higher spin algebra whose generators are constructed from
the product of generators $X_i$, $P_j$ up to second order:{\samepage
\begin{gather*}
\mathfrak{h}_2\subset{\rm Galilei}\subset{\rm Sch}(2)\subset\mathfrak{h}_2\oplus\mathfrak{sp}
(4)\subset[\mathfrak{h}_2^*].
\end{gather*}
${\rm Sch}(2)$ is the Schr\"odinger algebra\footnote{A f\/ield theory realization of this algebra was
given in~\cite{HMS03}.} in 2D, whose generators are those of the Galilean algebra $X_i$, $P_i$, $H$, $J$,
together with the dilatation, $D$, and the expansion, $C$.}

Let us restrict now to Galilean and Schr\"odinger symmetries.
We start by considering the Galilean symmetries of~\eqref{action}.
The action is invariant under translations,
\begin{gather*}
x_i^{\prime}=x_i+\alpha_i,
\qquad
v_i^{\prime}=v_i,
\end{gather*}
boosts,
\begin{gather*}
x_i^{\prime}=x_i-\beta_i t,
\qquad
v_i^{\prime}=v_i-\beta_i,
\end{gather*}
rotations,
\begin{gather*}
x'_i=x_i\cos\varphi+\epsilon_{ij}x_j\sin\varphi,
\qquad
v'_i=v_i\cos\varphi+\epsilon_{ij}v_j\sin\varphi,
\end{gather*}
and time translations
\begin{gather*}
t'=t-\gamma.
\end{gather*}
The corresponding Noether charges of translations and boosts are given by
\begin{gather*}
P_i=p_i,
\qquad
K_i=mx_i-p_it-\pi_i+\frac\kappa2\epsilon_{ij}v_j=mX_i+\frac{\kappa}{2m}\epsilon_{ij}P_j,
\end{gather*}
while the angular momentum is
\begin{gather}
\label{Jncir}
J=\epsilon_{ij}(x_ip_j+v_i\pi_j)=\epsilon_{ij}(X_iP_j+\tilde v_i\tilde\pi_j).
\end{gather}

Together with the total Hamiltonian~\eqref{quadratichamiltonian}, they generate the {exotic Galilei}
algebra~\cite{BGO,BGGK,HMS03,HMS10, LL}
\begin{gather*}
\{H,J\}=0,
\qquad
\{H,K_i\}=-P_i,
\qquad
\{H,P_i\}=0,
\qquad
\{J,P_i\}=\epsilon_{ij}P_j,
\\
\{J,K_i\}=\epsilon_{ij}K_j,
\qquad
\{K_i,P_j\}=m\delta_{ij},
\qquad
\{K_i,K_j\}=-\kappa\epsilon_{ij},
\qquad
\{P_i,P_j\}=0.
\end{gather*}
From this, it may seem that the Lagrangian~\eqref{action} gives a~phase space realization of the
$(2+1)$-dimensional Galilei group with two central charges $m$, $\kappa$.
However, one of the central charges is trivial since, if we modify the generator of the boost as
in~\cite{BGGK,Hagen:1984mj},
\begin{gather*}
\tilde K_i={K}_i-\frac{\kappa}{2m}\epsilon_{ij}{P}_j=m X_i=mx_i-\pi_i+\frac12\kappa\epsilon_{ij}
v_j-\frac{\kappa}{2m}\epsilon_{ij}p_j-p_it,
\end{gather*}
one gets that $(H,P,\tilde K,J)$ verif\/ies the {standard Galilean} algebra without $\kappa$.\footnote{If we
introduce an interaction with a~background f\/ield this statement is no longer true, since it depends on
which coordinates (commutative or non-commutative) are used to def\/ine the interaction;
see~\cite{delOlmo:2005md,Duval:2001hu, HMS10,Horvathy:2005wf}.
Notice however that the background f\/ield will break, in general, part of the symmetries of the Galilei
group.} Physically, the result of changing the boost generators is a~shift in the parameter of the
translations
\begin{gather*}
\alpha_i\to\alpha_i+\frac{\kappa}{2m}\epsilon_{ij}\beta_j.
\end{gather*}
Note that the modif\/ied boost generators $\tilde K_i$ are proportional to the coordinates at $t=0$,
$X_i= \tilde x^i(0)$, that verify $\{X_i, X_j\}=0$, and we have a~realization with only one non-trivial
central charge associated to the mass of the particle\footnote{Note however that $\delta_{K_i}
L=\delta_{\tilde K_i} L= \frac{{\rm d}}{{\rm d}t}(-mx_i-\frac\kappa2\epsilon_{ij}v_j)\beta_i$, where $\beta_i$ is boost
parameter.}.

The Schr\"odinger generators are those of the Galilean algebra $X_i$, $P_i$, $H$, $J$, and the dilata\-tion,~$D$, and the expansion, $C$, given by
\begin{gather*}
D=X_i P_i=x_ip_i-\frac{t}{m}p_i^2-\frac1m\pi_i p_i+\frac{\kappa}{2m}\epsilon_{ij}p_iv_j,
\\
C=m X_i X_i=m x_i^2+\frac1m t^2p_i^2+\frac1m\pi_i^2+\frac{\kappa^2}{4m}v_i^2+\frac{\kappa^2}{4m^3}
p_i^2-2t x_i p_i-2x_i\pi_i+\kappa\epsilon_{ij}x_i v_j
\\
\phantom{C=}
{}-\frac{\kappa}{m}\epsilon_{ij}x_i p_j+\frac{2}{m}
t p_i\pi_i
-\frac{\kappa}{m}t\epsilon_{ij}p_i v_j-\frac{\kappa}{m}\epsilon_{ij}\pi_i v_j+\frac{\kappa}{m^2}
\epsilon_{ij}\pi_i p_j-\frac{\kappa^2}{2m^2}v_i p_i.
\end{gather*}

In the same spirit, we also redef\/ine the generator of rotations as
\begin{gather*}
J=\epsilon_{ij}X_i P_j=\epsilon_{ij}x_i p_j-\frac{\kappa}{2m^2}p_i^2+\frac{\kappa}{2m}v_i p_i+\frac{1}{m}
\epsilon_{ij}p_i\pi_j,
\end{gather*}
which, up to square of constraints, coincides with~\eqref{Jncir}.

The new, non-zero Poisson brackets are
\begin{gather*}
\{D,C\}=-2{\rm}C,
\qquad
\{D,H\}=2{\rm}H,
\qquad
\{H,C\}=-2D,
\\
\{D,P_i\}={\rm}P_i,
\qquad
\{D,X_i\}=-{\rm}X_i,
\qquad
\{C,P_i\}=2{\rm}m X_i.
\end{gather*}

The transformations of the coordinates $x_i$, $v_i$ under dilatation and expansion are obtained
from~\eqref{delxvF} as
\begin{gather*}
\delta_{D}x_i=\frac{\alpha}{m}(m x_i-2mt v_i+\kappa\epsilon_{ij}v_j),
\qquad
\delta_{D}v_i=-\alpha  v_i,
\\
\delta_{C}x_i=\frac{\lambda}{m}\left(2mt^2v_i-2mtx_i+\kappa\epsilon_{ij}x_j-2\kappa t\epsilon_{ij}
v_j-\frac{\kappa^2}{2m}v_i\right),
\\
\delta_{C}v_i=\frac{\lambda}{m}(-2m x_i+2mt v_i-\kappa\epsilon_{ij}v_j),
\end{gather*}
where $\alpha$ and $ \lambda$ are the corresponding inf\/initesimal parameters.

\subsection{Reduction of second class constraints}\label{section2.1}

The classical symmetry algebra is also realized in the reduced phase space def\/ined by the second class
constraints $\Pi_i=V_i=0$.
The Dirac bracket is
\begin{gather*}
\{A,B\}^*=\{A,B\}+\{A,\Pi_i\}\frac1m\{V_i,B\}-\{A,V_i\}\frac1m\{\Pi_i,B\}-\{A,V_i\}\frac{\kappa\epsilon_{ij}
}{m^2}\{V_j,B\}
\end{gather*}
and yields
\begin{gather}
\label{xxpfree}
\{x_i,x_j\}^*=\frac{\kappa}{m^2}\epsilon_{ij},
\qquad
\{x_i,p_j\}^*=\delta_{ij},
\qquad
\{p_i,p_j\}^*=0.
\end{gather}

In this space, the symmetry transformations are generated using the Dirac bracket and the reduced
generators, which can be obtained by substituting $v_i=p_i/m$, $\pi_i=-\kappa/(2m)\epsilon_{ij}p_j$ into
the standard ones.

The inf\/inite Weyl symmetries are generated by
\begin{gather*}
\mathfrak{G}^{(R)}(x_i,p_j)=\mathfrak{G}(X_i,P_j)|_{v_i=p_i/m,\, \pi_i=-\kappa/(2m)\epsilon_{ij}p_j}.
\end{gather*}

In particular the Schr\"odinger generators are given by~\cite{Banerjee:2005zt}
\begin{gather}
P_i^{(R)}=p_i,
\label{red_class_genP}
\\
K_i^{(R)}=mx_i-tp_i+\frac{\kappa}{m}\epsilon_{ij}p_j
\qquad
\text{(exotic Galilei)},
\\
\tilde K_i^{(R)}=K_i^{(R)}-\frac{\kappa}{2m}\epsilon_{ij}P_j^{(R)}=mx_i-tp_i+\frac{\kappa}{2m}\epsilon_{ij}p_j
\qquad
\text{(standard Galilei)},
\\
H^{(R)}=\frac{1}{2m}p_i^2,
\\
J^{(R)}=\epsilon_{ij}x_i p_j+\frac{\kappa}{2m^2}p_i^2,
\\
D^{(R)}=p_ix_i-\frac{1}{m}t p_i^2,
\\
C^{(R)}=mx_i^2+\frac{1}{m}t^2p_i^2+\frac{\kappa^2}{4m^3}p_i^2-2t x_i p_i+\frac{\kappa}{m}\epsilon_{ij}
x_i p_j.
\label{red_class_genC}
\end{gather}
They generate the Schr\"odinger algebra with the Dirac bracket, since $\tilde K_i^{(R)}$, $P_i^{(R)}$ generate
a~Heisenberg algebra:
\begin{gather*}
\left\{\tilde K_i^{(R)},P_j^{(R)}\right\}^*=m\delta_{ij},
\qquad
\left\{P_i^{(R)},P_j^{(R)}\right\}^*=0,
\qquad
\text{and}
\qquad
\left\{\tilde K_i^{(R)},\tilde K_j^{(R)}\right\}^*=0.
\end{gather*}

Symmetry transformations are generated either using the Poisson brackets in the original phase space or
using the Dirac brackets with the reduced generators,~\eqref{red_class_genP}--\eqref{red_class_genC}.
For example the ``exotic Galilei" generators $K_i$ satisfy
\begin{gather*}
\left\{K_i,K_j\right\}=\left\{K_i^{(R)},K_j^{(R)}\right\}^*=-\kappa\epsilon_{ij},
\end{gather*}
and generate ``standard(covariant) Galilei" transformation of $(x_i,p_i)$ as
\begin{gather*}
\delta x_i=\{x_i, \beta\cdot K\}=\big\{x_i, \beta\cdot K^{(R)}\big\}^*=-t\beta_i,
\qquad
\\
\delta p_i=\{p_i, \beta\cdot K\}=\big\{p_i, \beta\cdot K^{(R)}\big\}^*=-m\beta_i.
\end{gather*}
The ``standard Galilei" generators $\tilde K_i$ satisfy
\begin{gather*}
\big\{\tilde K_i,\tilde K_j\big\}=\left\{\tilde K_i^{(R)},\tilde K_j^{(R)}\right\}^*=0.
\end{gather*}
and generate ``exotic Galilei'' (non-covariant) transformations of $x_i$, $p_i$,
\begin{gather*}
\delta x_i=\{x_i, \beta\cdot\tilde K\}=\big\{x_i, \beta\cdot\tilde K^{(R)}\big\}^*=-t\beta_i+\frac{\kappa}{2m}
\epsilon_{ij}\beta_j,
\qquad
\\
\delta p_i=\{p_i, \beta\cdot\tilde K\}=\big\{p_i, \beta\cdot\tilde K^{(R)}\big\}^*=-m\beta_i.
\end{gather*}

\section{Quantum symmetries of free Schr\"odinger equation\\
in the non-commutative plane}
\label{section3}

In this section we will study the quantization of the model {at the level of} the Schr\"odinger equation
and their symmetries.
We will quantize it in {two} approaches, one in the reduced phase space and the other in the extended phase
space.

\subsection{Quantization in the reduced phase space}

In the classical theory, $x_i$ has a~nonzero Dirac bracket $\{x_i,x_j\}^*$ as in~\eqref{xxpfree} in the
reduced phase space.
Since Dirac brackets are replaced by commutators in the canonical quantization, one cannot have
a~$x_i$-coordinate representation of quantum states\footnote{Since $p_i$'s are commuting the momentum
representation is possible~\cite{Duval:2001hu}.}.
To discuss symmetries of Schr\"odinger equations we introduce new coordinates
\begin{gather}
\label{def_comm-coor0}
y_i\equiv x_i+\frac{\kappa}{2m^2}\epsilon_{ij}p_j,
\qquad
q_i=p_i,
\end{gather}
such that
\begin{gather*}
\{y_i,y_j\}^*=0,
\qquad
\{y_i,q_j\}^*=\delta_{ij},
\qquad
\{q_i,q_j\}^*=0.
\end{gather*}
The coordinate $y_i$ is the one introduced in~\eqref{def_comm-coor1} and $q_i$ is its conjugate.
In these coordinates, the Schr\"odinger equation $(\text{i}{\partial_t}-H) |\Psi(t)\rangle=0$ takes the form
corresponding to a~free particle for the wave function
\begin{gather*}
\Psi(y,t)=\langle y|\Psi(t)\rangle,
\qquad
\hat y_i|y\rangle=y_i|y\rangle,
\qquad
\langle y|y'\rangle=\delta^2(y-y'),
\end{gather*}
i.e.               
\begin{gather*}
\left(\text{i}{\partial_t}-\frac1{2m}(-\text{i}\partial_y)^2\right)\Psi\left(y,t\right)=0,
\end{gather*}
and the inner product is
\begin{gather*}
\langle\Psi|\Psi\rangle=\int {\rm d}y\,{\ov{{\Psi\left(y,t\right)}}}\Psi\left(y,t\right).
\end{gather*}
Note that $y_i$ are not covariant under exotic Galilei transformation generated by $K_i$
\begin{gather*}
\delta y_i=\{y_i, \beta\cdot K\}=\big\{y_i, \beta\cdot K^{(R)}\big\}^*=-\beta_i t-\frac{\kappa}{2m}\epsilon_{ij}
\beta_j,
\end{gather*}
but covariant under the Galilei transformation generated by $\tilde K_i$
\begin{gather*}
\delta y_i=\{y_i, \beta\cdot\tilde K\}=\big\{y_i, \beta\cdot\tilde K^{(R)}\big\}^*=-\beta_i t.
\end{gather*}
The position operators, covariant under $K_i$, are
\begin{gather*}
\hat x_i=y_i-\frac{\kappa}{2m^2}\epsilon_{ij}(-\text{i}\partial_{y_j}).
\end{gather*}
They are hermitian since $\hat y_i=y_i$, $\hat q_i=-{\text{i}}\partial_{y_i}$, with appropriate
boundary conditions on $\Psi \left( y,t\right)$, are hermitian.

Although in the free theory we are able to work with both the commutative $\hat y_i=y_i$ and the
non-commutative $\hat x_i=y_i-\frac{\kappa}{2m^2}\epsilon_{ij}(-\text{i}\partial_{y_j})$ position operators, this may
not be the case in an interacting theory.
For example, if we consider an interaction with a~background electromagnetic f\/ield, which introduces
couplings with a~source at position $x_i$, the non-commutative coordinates are naturally selected (see, for
example,~\cite{delOlmo:2005md,Duval:2001hu, HMS10,Horvathy:2005wf}).
If we denote generically by $\mathfrak{G}^{(R)}(t,x,p)=\mathfrak{G}(X,P)|_{\Pi=V=0}$ the generators of the
{Weyl} algebra in the reduced classical space, the generators in this quantization are given by{\samepage
\begin{gather}
\label{G(1)}
\hat{\mathfrak{G}}_i^{(1)}(t,y,\hat q)=\left.
\mathfrak{G}_i^{(R)}\right|_{x_j=y_j-\frac{\kappa}{2m^2}\epsilon_{jl}\hat q_l,\, p_j=\hat q_j}=\mathfrak{G}_i\left(y-\frac t m{\hat q},\hat q\right),
\end{gather}
with $\hat q_i=-\text{i}\derpar{}{y_i}$ and with the appropriate dealing of operator ordering.}

The knowledge of all the symmetries of the Schr\"odinger equation in terms of the coordina\-tes~$y^i$,~$\hat y_i$ is the non-commutative analog in $2+1$ dimensions of the high spin symmetries of the relativistic
massless Klein Gordon equation~\cite{Eastwood:2002su}.
The Vasiliev~\cite{Vasiliev:2004cp} non-linear theory has these high spin symmetries.
In this sense these high spin-nonrelativistic symmetries could be useful in order to construct
a~non-relativistic Vasiliev theory~\cite{Valenzuela:2009gu}.

We consider next in detail the Schr\"odinger generators, given by
\begin{gather}
\hat P_i^{(1)}=\hat q_i=-{\text{i}}\derpar{}{y_i},
\\
\hat{\tilde K}_i^{(1)}=m y_i-t\hat q_i=m y_i+{\text{i}}t\derpar{}{y_i},
\\
\hat H^{(1)}=\frac{1}{2m}\hat q_i^2=-\frac{1}{2m}\frac{\partial^2}{\partial{y_i}^2},
\\
\hat J^{(1)}=\epsilon_{ij}y_i\hat q_j=-{\text{i}}\epsilon_{ij}y_i\derpar{}{y_j},
\\
\hat D^{(1)}=y_i\hat q_i-{\text{i}}-\frac{1}{m}t\hat q_i^2=-{\text{i}}y_i\derpar{}
{y_i}+\frac{1}{m}t\frac{\partial^2}{\partial{y_i}^2}-{\text{i}},
\\
\hat C^{(1)}=m y_i^2-2t y_i\hat q_i+2{\text{i}}t+\frac{1}{m}
t^2\hat q_i^2=m y_i^2+2{\text{i}}t y_i\derpar{}{y_i}-\frac{1}{m}t^2\frac{\partial^2}
{\partial{y_i}^2}+2{\text{i}}t,
\label{red_quantum_genC}
\end{gather}
where a~Weyl ordering has been used for $\hat D^{(1)}$ and $\hat C^{(1)}$.
These generators are hermitian operators when acting on the wave functions $\Psi(t,y)$.
Furthermore, they obey the abstract quantum Schr\"odinger algebra \textit{off shell}, with non-zero
commutators given by
\begin{gather}
\big[\hat{\tilde K}_i,\hat P_j\big]={\text{i}}m\delta_{ij},
\qquad
\big[\hat J,\hat P_i\big]={\text{i}}\epsilon_{ij}\hat P_j,
\qquad
\big[\hat J,\hat{\tilde K}_i\big]={\text{i}}\epsilon_{ij}\hat{\tilde K}_j,
\qquad
\big[\hat H,\hat{\tilde K}_i\big]=-{\text{i}}\hat P_i,\nonumber
\\
\big[\hat D,\hat H\big]=2{\text{i}}\hat H,
\qquad
\big[\hat D,\hat P_i\big]={\text{i}}\hat P_i,
\qquad
\big[\hat D,\hat{\tilde K}_i\big]=-{\text{i}}\hat{\tilde K}_i,\nonumber
\\
\big[\hat D,\hat C\big]=-2{\text{i}}\hat C,
\qquad
\big[\hat H,\hat C\big]=-2{\text{i}}\hat D,
\qquad
\big[\hat C,\hat P_i\big]=2{\text{i}}\hat{\tilde K}_i.
\label{carles10}
\end{gather}
Using these, together with
\begin{gather}
\big[{\text{i}}\partial_t,\hat{\tilde K}_i^{(1)}\big]=-{\text{i}}\hat P_i^{(1)},
\qquad
\big[{\text{i}}\partial_t,\hat D^{(1)}\big]=-2{\text{i}}\hat H^{(1)},
\qquad
\big[{\text{i}}\partial_t,\hat C^{(1)}\big]=-2{\text{i}}\hat D^{(1)},
\label{PKSch}
\end{gather}
one can show that
\begin{gather*}
\left[{\text{i}}\partial_t-\hat H^{(1)},\hat{\mathfrak{G}}_i^{(1)}\right]=0
\end{gather*}
for all the generators $\hat{\mathfrak{G}}_i^{(1)}$, which proves the invariance of the Schr\"odinger
equation under the Schr\"odinger transformations in this reduced space quantization.

Under a~general Weyl transformation, the wave functions transform as
\begin{gather*}
\Psi'\left(y,t\right)=e^{\text{i}\alpha_i\hat{\mathfrak{G}}_i^{(1)}(t,y,(-\text{i}\partial_y))}\Psi\left(y,t\right),
\end{gather*}
where the $\alpha_i$ are the parameters of the transformations.
In particular, for the {\it on-shell} Schr\"o\-din\-ger transformations one has
\begin{gather*}
\Psi'\left(y,t\right)=e^{A+iB}\Psi\left(y',t'\right),
\end{gather*}
where the coordinate transformations of $(y,t)$ are those of the $N=1$ conformal Galilean transformation,
and the multiplicative factor is $e^{A+iB}$, with $A$ and $B$ real functions of the coordinates and of the
parameters of the transformation given by (see, for instance,~\cite{Gomis:2011dw, Niederer:1972zz})
\begin{enumerate}\itemsep=-1pt
\item[1)] $H$ (time translation),
\begin{gather*}
t'=t+a,
\qquad
y'=y,
\qquad
A=B=0,
\end{gather*}
\item[2)] $D$ (dilatation),
\begin{gather*}
t'=e^\lambda t,
\qquad
y'=e^{\frac\lambda2}y,
\qquad
A=\frac{\lambda}2,
\qquad
B=0,
\end{gather*}
\item[3)] $C$ (expansion),
\begin{gather*}
t'=\frac{t}{1-\kappa t},
\qquad
y'_i=\frac{y_i}{1-\kappa t},
\qquad
e^A=\frac{1}{(1-\kappa t)},
\qquad
B=-\frac{\kappa m y^2}{2(1-\kappa t)},
\end{gather*}
\item[4)] (spatial translations and boost)
\begin{gather*}
t'=t,
\quad
y'_i=y_i+\left(\beta^0+t\frac{\beta^1}{m}\right)_i,
\qquad
A=0,
\quad
B=-m\left(y_i+\frac12\left(\beta_i^0+t\frac{\beta_i^1}{m}\right)\right)\frac{\beta_i^1}{m},
\end{gather*}
with $[\beta_i^0]=L,\;[\beta_i^1]=L^{-1} $.
\end{enumerate}

The dif\/ference with respect to the transformation of the ordinary Schr\"odinger equation is that in the
non-commutative case the coordinates that are transformed by conformal Galilean transformations are the
canonical ones~$y_i$, and not the physical position of the particle, $x_i$.

The invariance of the solutions of the Schr\"odinger equation under a~general element of the Weyl algebra
can be proved using the invariance under the generators of the Heisenberg algebra and the
commutators~\eqref{PKSch}.

\subsection{Quantization in the extended phase space}

\subsubsection{Fock representation}

In order to quantize the model in the extended phase space the second class constraints~\eqref{constNon2}
are imposed as physical state conditions by taking their non-hermitian combinations as
in~\cite{Alvarez:2007fw}.
We f\/irst consider the canonical transformation~\eqref{canonicaltransformation} that separates the second
class constraints as new coordinates.
It is realized at quantum level as a~unitary transformation
\begin{gather}
\tilde q=U^\dagger q U,
\qquad
U=e^{\frac{\text{i}}{m}p_i(\pi_i-\frac\kappa2\epsilon_{ij}v_j)}.
\label{U}
\end{gather}
For example,
\begin{gather*}
\tilde x_i=U^\dagger x_iU=x_i-\frac1m\left(\pi_i-\frac\kappa2\epsilon_{ij}v_j\right)+\frac12\frac{\kappa}{m}\epsilon_{ij}
\left(-\frac{p_j}{m}\right).
\end{gather*}

It is useful to introduce the complex combinations of the phase space variables
$\tilde\pi_\pm=\tilde\pi_1\pm{\text{i}}\tilde\pi_2$ and $\tilde v_\pm=\tilde v_1\pm{\text{i}}\tilde v_2$,
which allow us to introduce two pairs of annihilation and creation operators
\begin{gather*}
{\tilde a}_\pm=\frac{\text{i}}{\sqrt{2\kappa}}\left(\tilde\pi_\pm-\text{i}\frac{\kappa}2\tilde v_\pm\right),
\qquad
{\tilde a}_\pm^\dagger=\frac{-\text{i}}{\sqrt{2\kappa}}\left(\tilde\pi_\mp+\text{i}\frac{\kappa}2\tilde v_\mp\right),
\qquad
\end{gather*}
with nonzero commutators $ [{\tilde a}_\pm,{\tilde a}^\dagger_\pm]=1 $.
Using the Fock representation for $(\tilde v,\tilde\pi)$ and coordinate representation for $(\tilde x, \tilde p)$, any
state of this system is described by
\begin{gather*}
|\Psi(t)\rangle=\sum_{n_+\geq0,n_-\geq0}\int {\rm d}y\, |n_+,n_-\rangle\otimes|y\rangle \Phi_{n_+n_-}(y,t),
\end{gather*}
where $ |n_+,n_-\rangle$ is the eigenstate of $\tilde N_\pm=\tilde a^\dagger_\pm \tilde a_\pm$ with eigenvalues $n_\pm\in
\mathbb{N}\cup\{0\}$ and $ |y\rangle$ is the eigenstate of commuting operators $\tilde x_i$ with eigenvalue
$y_i$.
They are normalized as
\begin{gather*}
\langle n_+,n_-|n_+',n_-'\rangle=\delta_{n_+n_+'}\delta_{n_-n_-'},
\qquad
\langle y|y'\rangle=\delta^2(y-y').
\end{gather*}
The scalar product is given by
\begin{gather*}
\langle\Psi|\Psi'\rangle=\sum_{n_\pm}\int {\rm d}y\,{\overline{\Phi_{n_+n_-}(y,t)}}\Phi'_{n_+n_-}(y,t).
\end{gather*}

In the quantization in the extended phase space the second class constraints~\eqref{constNon2} are imposed
as physical state conditions by taking their non-hermitian combination,
\begin{gather}
\tilde a_\pm|\Psi_{\rm phys}(t)\rangle=0.
\label{7apmphys}
\end{gather}
This means that physical states are minimum uncertainty states in $(\tilde v,\tilde\pi)$.
Condition~\eqref{7apmphys} selects out only the ${n_+=n_-}=0$ state, so that $\Phi_{n_+n_-}(y,t)=0$ except
for $\Phi_{0,0}(y,t)\equiv \Phi_{0}(y,t)$,
\begin{gather*}
|\Psi_{\rm phys}(t)\rangle=\int {\rm d}y\,|0,0\rangle\otimes|y\rangle \Phi_0(y,t).
\end{gather*}
The Schr\"odinger equation is
\begin{gather*}
(\text{i}\partial_t-H)|\Psi_{\rm phys}(t)\rangle=0,
\qquad
H=\frac{\hat{\tilde p}^2}{2m},
\end{gather*}
and thus
\begin{gather*}
(\text{i}\partial_t-H){\Phi}_0(y,t)=0,
\qquad
H=\frac1{2m}(-\text{i}\partial_{y_i})^2.
\end{gather*}

The generators of the Weyl algebra are given in the extended space as polynomials $\mathfrak{G}(X,P)$ of
the operator equivalent of~\eqref{XPdef}, and, since they commute with $\tilde a_\pm$ and $ \tilde a^\dagger_\pm$,
physical states remain physical\footnote{The angular momentum $J$ in~\eqref{Jncir} contains a~term
depending on $(v,\pi)$, but it commutes with $\tilde a_\pm$, $\tilde a^\dagger_\pm$.}.
They act on the physical states as
\begin{gather*}
\Psi_{\rm phys}(t)\rangle\quad\to\quad|\Psi'_{\rm phys}(t)\rangle=e^{\text{i}\mathfrak{G}(X,P)}|\Psi_{\rm phys}
(t)\rangle
\end{gather*}
and it turns out that the transformation of the wave function $\Phi_0(y,t)$ is
\begin{gather*}
\Phi'_{0}(y,t)=e^{\text{i}\mathfrak{G}(X,P)}\Phi_{0}(y,t)=e^{\text{i}\mathfrak{G}(y-t(-\text{i}\partial_y),(-\text{i}\partial_y))}
\Phi_0(y,t).
\end{gather*}
This transformation has the same form as the one in the reduced phase space generated
by \linebreak \mbox{\eqref{G(1)}--\eqref{red_quantum_genC}}.
Then the wave function in the reduced space $ {\Psi}(y,t)=\langle y |\Psi(t)\rangle$ and $
{\Phi}_0(y,t)=\langle y|\otimes \langle00|\Psi(t)\rangle$ that appear in the extended space quantization
are identif\/ied.
Note that in the former~$\langle y |$ is eigenstate of $\hat y_i=x_i+\frac{\kappa}{2m^2}\epsilon_{ij}p_j$
in~\eqref{def_comm-coor0} but $\langle y |$ in the latter is eigenstate of $\hat{\tilde x}_i$ that are commuting
in the extended space.

We can see now how the non-commutativity of the position operators appears.
$\hat x_\pm=x_1\pm \text{i}x_2$ are commuting in the extended phase space.
Using~\eqref{canonicaltransformation} we write
\begin{gather*}
x_+=\tilde x_++\text{i}\frac{\kappa}{2m^2}\tilde p_++\text{i}\sqrt{\frac{2\kappa}{m^2}}\tilde a_-^\dagger,
\qquad
x_-=\tilde x_--\text{i}\frac{\kappa}{2m^2}\tilde p_--\text{i}\sqrt{\frac{2\kappa}{m^2}}\tilde a_-=x_+^\dagger.
\end{gather*}
In the reduced space quantization procedure, the $\tilde a_\pm$ are ef\/fectively put to zero and $x_\pm$
becomes a~non-commutative operator on~$|\Psi(t)\rangle$.
On the other hand in the quantization in the extended space, expectation values of the position operators
between two physical states are given by
\begin{gather*}
\begin{split}
& \langle\Psi|\hat x_\pm|\Psi'\rangle=\int {\rm d}y{\rm d}y'\,{\overline{\Phi_0(y,t)}}
\langle y|\langle0|\left(\tilde x_\pm\pm \text{i}\frac{\kappa}{2m^2}\tilde p_\pm\pm \text{i}\sqrt{\frac{2\kappa}{m^2}}
\begin{pmatrix}
\tilde a_-^\dagger\cr\tilde a_-
\end{pmatrix}
\right)|0\rangle|y'\rangle \Phi'_{0}(y',t)
\\
& \phantom{\langle\Psi|\hat x_\pm|\Psi'\rangle}
=\int {\rm d}y\,{\overline{\Phi_0(y,t)}}\left(y_\pm\pm \text{i}\frac{\kappa}{2m^2}(-2\text{i}\partial_{y_\pm})\right)\Phi'_0(y,t).
\end{split}
\end{gather*}
Commutative position operators $\hat x_\pm$ on states $|\Psi\rangle$ act as non-commutative operators $(y_\pm
\pm \text{i}\frac{\kappa}{2m^2} (-2\text{i}\partial_{y_\pm}))$ on the wave function $\Phi_0(y,t)$.

It is useful to consider the unitary transformation $U$ in~\eqref{U} on the creation and annihilation
operators $\tilde a_\pm$, $\tilde a^\dagger_\pm$,
\begin{gather}
\label{condition2}
\tilde a_+=U^\dagger a_+U=a_+,
\qquad
\tilde a_-=U^\dagger a_-U=a_--\sqrt{\frac\kappa{2m^2}}p_-.
\end{gather}
The quantization in the extended phase space can be also done by considering the constraint
equations~\eqref{7apmphys} in terms of the operators $a_\pm$, $a^\dagger_\pm$.
The physical state conditions~\eqref{7apmphys} are
\begin{gather*}
a_+|\Psi_{\rm phys}(t)\rangle=0,
\qquad
\left(p_--\sqrt{\frac{2m^2}{\kappa}}a_-\right)|\Psi_{\rm phys}(t)\rangle=0,
\end{gather*}
and $|\Psi_{\rm phys}\rangle$ is a~coherent state of $a_-$ with eigenvalue
$\sqrt{\frac\kappa{2m^2}}p_-$~\cite{Horvathy:2004fw}.
In this representation, the Schr\"odinger generators are
\begin{gather*}
X_\pm^{(2)}=\left(x_\pm\mp \text{i}\frac{\kappa}{2m^2}p_\pm\right)-\frac{t}m p_\pm\pm \text{i}\frac{\kappa}{m^2}
\left(p_\pm-\sqrt{\frac{2m^2}\kappa}
\begin{pmatrix}
a_-^\dagger\cr a_-
\end{pmatrix}
\right),
\\
P_\pm^{(2)}=p_\pm=-{2} \text{i}\partial_{x_\mp},
\qquad
[x_{\pm},p_{\mp}]={2} \text{i},
\\
D^{(2)}=\frac{1}2\Bigg(\left(x_+p_-+p_+x_--\frac{2t}{m}p_+p_-\right)
+\text{i}\frac{\kappa}{m^2}\left(p_+-\sqrt{\frac{2m^2}\kappa}{a^\dagger_-}\right)p_-
\\
\phantom{D^{(2)}=}
{}-\text{i}\frac{\kappa}{m^2}p_+\left(p_--\sqrt{\frac{2m^2}\kappa}{a_-}\right)\Bigg),
\\
C^{(2)}=\frac{1}2\Bigg(\left(x_+-\text{i}\frac{\kappa}{2m^2}p_+\right)\left(x_-+\text{i}\frac{\kappa}{2m^2}p_-\right)
\\
\phantom{D^{(2)}=}
{}-\frac{t}m\left(\left(x_+-\text{i}\frac{\kappa}{2m^2}p_+\right)p_-+p_+\left(x_-+\text{i}\frac{\kappa}{2m^2}p_-\right)\right)
\\
\phantom{C^{(2)}=}
{}+\frac{t^2}{2m^2}p_+p_-+\frac{1}2\left(\left(x_+-\text{i}\frac{\kappa}{2m^2}p_+\right)-\frac{t}m p_+\right)
\left(-\text{i}\frac{\kappa}{m^2}\right)\left(p_--\sqrt{\frac{2m^2}\kappa}{a_-}\right)
\\
\phantom{C^{(2)}=}
{}+\frac{1}2\text{i}\frac{\kappa}{m^2}\left(p_+-\sqrt{\frac{2m^2}\kappa}{a_-^\dagger}\right)
\left(\left(x_-+\text{i}\frac{\kappa}{2m^2} p_-\right)-\frac{t}m p_-\right)
\\
\phantom{C^{(2)}=}
{}+\frac12\left(\text{i}\frac{\kappa}{m^2}\left(p_+-\sqrt{\frac{2m^2}\kappa}{a_-^\dagger}\right)\right)
\left(-\text{i}\frac{\kappa}{m^2}\left(p_--\sqrt{\frac{2m^2}\kappa}{a_-}\right)\right)\Bigg).
\\
J^{(2)}=\frac{\text{i}}2\Bigg(\left(x_+p_--p_+x_--\text{i}\frac{\kappa}{m^2}p_+p_-\right)
+\text{i}\frac{\kappa}{m^2}\left(p_+-\sqrt{\frac{2m^2}\kappa}{a^\dagger_-}\right)p_-
\\
\phantom{J^{(2)}=}
{}+\text{i}\frac{\kappa}{m^2}p_+\left(p_--\sqrt{\frac{2m^2}\kappa}{a_-}\right)\Bigg).
\end{gather*}

These generators commute with the constraint equations and with the Schr\"odinger operator $\text{i}\partial_t-H$.
Notice that the set of generators do not depend on $a_+$, $a^\dagger_+$, and therefore the transition to
the Fock space used in~\cite{Horvathy:2004fw} is recovered.

The Fock expression of a~generic element of the Weyl algebra $\mathfrak{G}(X,P)$ can be obtained using the
expression of the operators $X$ and $P$ given by~\eqref{XPdef}.

\subsubsection{Coordinate representation}

In the representation of coordinates the time Schr\"odinger equation and the constraint
equations~\eqref{condition2} in the non-commutative plane becomes~\cite{Alvarez:2007fw}
\begin{gather*}
\hat S_1\Psi\equiv\left(\frac{\partial}{\partial v_{-}}+\frac{\kappa}{4}v_{+}
\right)\Psi\left(x,v,t\right)=0,
\\
\hat S_2\Psi\equiv\left(\frac{\partial}{\partial x_{+}}-{\text{i}}\frac{m}{4}v_{-}
-{\text{i}}\frac{m}{\kappa}\frac{\partial}{\partial v_{+}}\right)\Psi\left(x,v,t\right)=0,
\\
\hat S_3\Psi\equiv\left({\text{i}}\frac{\partial}{\partial t}+\frac{2}{m}\frac{\partial^{2}}
{\partial x_{+}\partial x_{-}}\right)\Psi\left(x,v,t\right)=0.
\end{gather*}
In this representation, the operators associated to the generators of the Heisenberg algebra are
\begin{gather*}
\hat P_1=-{\text{i}}\derpar{}{x_+}-{\text{i}}\derpar{}{x_-},
\qquad
\hat P_2=\derpar{}{x_+}-\derpar{}{x_-},
\\
\hat{\tilde K}_1=\frac{m}{2}(x_++x_-)+\left({\text{i}}t-\frac{\kappa}{2m}\right)\derpar{}{x_+}
+\left({\text{i}}t+\frac{\kappa}{2m}\right)\derpar{}{x_-}+\frac{\kappa}{4{\text{i}}}
(v_+-v_-)+{\text{i}}\derpar{}{v_+}+{\text{i}}\derpar{}{v_-},
\\
\hat{\tilde K}_2=\frac{m}{2{\text{i}}}(x_+-x_-)-\left(t+{\text{i}}\frac{\kappa}{2m}
\right)\derpar{}{x_+}+\left(t-{\text{i}}\frac{\kappa}{2m}\right)\derpar{}{x_-}-\frac{\kappa}{4}
(v_++v_-)-\derpar{}{v_+}+\derpar{}{v_-},
\end{gather*}
or, in covariant form,
\begin{gather*}
\hat P_i=-{\text{i}}\derpar{}{x_i},
\qquad
\hat{\tilde K}_i=m x_i+{\text{i}}t\derpar{}{x_i}+{\text{i}}\frac{\kappa}{2m}
\epsilon_{ij}\derpar{}{x_j}+\frac{\kappa}{2}\epsilon_{ij}v_j+{\text{i}}\derpar{}{v_i},
\end{gather*}
which, indeed, satisfy $[\hat P_i,\hat {\tilde K}_j] = -{\text{i}} m \delta_{ij}$, with all the
other commutators equal to zero.

It is immediate to check that the operators $\hat P_i$, $\hat{\tilde K}_i$ commute with all of $\hat S_1$,
$\hat S_2$ and $\hat S_3$, and hence that they generate Schr\"odinger symmetries for the free particle in
the non-commutative plane.
The rest of generators of the Schr\"odinger algebra are given by
\begin{gather*}
\hat H=-\frac{2}{m}\dderpar{}{x_+}{x_-}=-\frac{1}{2m}\frac{\partial^2}{\partial{x_i}^2},
\\
\hat J=-{\text{i}}\epsilon_{ij}x_i\derpar{}{x_j}+\frac{\kappa}{2m^2}\frac{\partial^2}
{\partial{x_i}^2}-{\text{i}}\frac{\kappa}{2m}v_i\derpar{}{x_i}-\frac{1}{m}\epsilon_{ij}\dderpar{}{x_i}{v_j},
\\
\hat D=-{\text{i}}x_i\derpar{}{x_i}+\frac{1}{m}t\frac{\partial^2}{\partial{x_i}^2}+\frac{1}{m}
\dderpar{}{x_i}{v_i}+{\text{i}}\frac{\kappa}{2m}\epsilon_{ij}v_i\derpar{}{x_j}-{\text{i}},
\\
\hat C=2{\text{i}}t x_i\derpar{}{x_i}+{\text{i}}\frac{\kappa^2}{2m^2}v_i\derpar{}
{x_i}+{\text{i}}\frac{\kappa}{m}\epsilon_{ij}x_i\derpar{}{x_j}-{\text{i}}\frac{\kappa}
{m}t\epsilon_{ij}v_i\derpar{}{x_j}
\\
\phantom{\hat C=}
{}-{\text{i}}\frac{\kappa}{m}\epsilon_{ij}v_i\derpar{}{v_j}+2{\text{i}}x_i\derpar{}
{v_i}-\frac{1}{m}t^2\frac{\partial^2}{\partial{x_i}^2}
-\frac{\kappa^2}{4m^3}\frac{\partial^2}{\partial{x_i}^2}-\frac{1}{m}\frac{\partial^2}{\partial{v_i}^2}
\\
\phantom{\hat C=}
{}-\frac{2}{m}t\dderpar{}{x_i}{v_i}+\frac{\kappa}{m^2}\epsilon_{ij}\dderpar{}{x_i}{v_j}
+m x_i^2+\kappa\epsilon_{ij}x_i v_j+\frac{\kappa^2}{4m}v_i^2+2{\text{i}}t.
\end{gather*}

Using these expressions, one can check explicitly the commutators~\eqref{carles10}, and also that these
quadratic generators commute with $\hat S_1$, $\hat S_2$ and $\hat S_3$ (this also follows from the
derivation properties of the commutators and the corresponding commutation of the linear generators $\hat
P_i$, $\hat{\tilde K}_i$, and this proves that the Schr\"odinger equation for the free particle in the
noncommutative plane has the Schr\"odinger algebra as a~symmetry.
Notice, however, that in this coordinate representation of the non-reduced quantum space the quadratic
operators contain second order derivatives, and hence do not generate point transformations for the
coordinates $x$, $v$.
This is in agreement with the results obtained in the reduced space quantization and the Fock space
representation.
In any case, the fact that the linear generators commute with $\hat S_1$, $\hat S_2$ and $\hat S_3$ allows
to prove that the quadratic ones also commute, and thus generate symmetries of the Schr\"odinger equation
of the free particle in the non-commutative plane.

\subsection*{Acknowledgments}

We thank Jorge Zanelli for collaboration in some parts of this work and
Mikhail Plyushchay for reading the manuscript.
We also thank Adolfo Azc\'arraga and Jurek Lukierski for discussions, and Rabin Banerjee for letting us
know about the results in~\cite{Banerjee:2005zt}.
CB was partially supported by Spanish Ministry of Economy and Competitiveness project DPI2011-25649.
We also acknow\-ledge partial f\/inancial support from projects FP2010-20807-C02-01, 2009SGR502 and CPAN
Consolider CSD 2007-00042.

\pdfbookmark[1]{References}{ref}
\LastPageEnding

\end{document}